\documentclass[aps,pra,twocolumn,groupedaddress]{revtex4-1}
\usepackage{units}
\usepackage{amsmath}
\usepackage{amssymb}
\usepackage{graphicx}
\usepackage{bm}
\usepackage{multirow,color,relsize,microtype}
\usepackage{dcolumn}
\usepackage{xcolor}

\newcommand{\be}{\begin{equation}}
\newcommand{\ee}{\end{equation}}
\newcommand{\pt}{{\cal PT}}

\begin{document}

\title{Time-reversal invariant scaling of light propagation in one-dimensional non-Hermitian systems}

\author{Jose D. H. Rivero}
\author{Li Ge}
\affiliation{Department of Physics and Astronomy, College of Staten Island, CUNY, Staten Island, New York 10314, USA}

\date{\today}

\begin{abstract}
Light propagation through a normal medium is determined not only by the real part of the refractive index but also by its imaginary part, which represents optical gain and loss. Therefore, two media with different gain and loss landscapes can have very different transmission and reflection spectra, even when their real parts of the refractive index are identical. Here we show that while this observation is true for an arbitrary one-dimensional medium with refractive index $n(x)$ and its time-reversed partner with refractive index $n^*(x)$, there exists a universal scaling that gives identical transmittance and reflectance in these corresponding systems. Interestingly, these scaled transmittance and reflectance reduce to their standard, unscaled forms in a time-reversal invariant system, i.e., one without gain or loss.

\end{abstract}


\maketitle
\section{Introduction}

Recent advances in nanofabrication and integration of photonic devices have profound technological impact on computation, communication and sensing \cite{tatebayashi_room-temperature_2015,computer,yu_complete_2009,LEE2009209}. They rely on the transport of information through optical structures, which has made the study of wave transport, localization and resonances essential to engineer their properties on demand. The scattering matrix is one of the central objects in this field of study \cite{Genack1987,Newton1982,Wiersma1997,Mostafazadeh2009,moiseyev,Mello2004,Ge2017}. Besides its wide range of applicability in optics and photonics, the scattering matrix has been used oftentimes to understand resonances in nuclear and particle physics \cite{particles1,particles2}, transport in condensed matter \cite{condensed} and in general, to probe states of open quantum systems \cite{rotter_NH}. 

The scattering matrix connects the incoming channels to the outgoing channels in a system, and in one-dimensional (1D) systems it consists of the transmission and reflection coefficients from both sides of the system. These quantities in a normal medium are determined not only by the real part of the refractive index but also by its imaginary part, which represents optical gain and loss. Such non-Hermitian systems \cite{NPreview} have attracted enormous interest in optics because of unique emerging phenomena such as spontaneous symmetry breaking \cite{NPhysreview}, coherent perfect absorption \cite{longhiCPA,CPA,CPA_exp,CPALaser}, anisotropic transmission resonances \cite{Ge2012,lin_unidirectional_2011,miao_orbital_2016}, self-sustained radiation \cite{schomerus}, asymmetric power oscillations \cite{PowerOsc,ge_flux_2017}, among others.

In general, scattering information contained in the reflection and transmission coefficients for a given frequency is different for waves in a gain or a loss system. Therefore, two media with different gain and loss landscapes can have very different transmission and reflection spectra, even when their real parts of the refractive index are identical. For example, consider a single slab of uniform refractive index. The transmission and reflection can become one order of magnitude larger when we add gain to overcompensate the intrinsic material loss. Furthermore, the frequency dependence of the transmission and reflection spectra can also become very different once the loss or gain landscape is changed, especially in more complicated systems.

Here we show that while these observations are true for an arbitrary 1D non-Hermitian medium with refractive index $n(x)$ and its time-reversed partner with refractive index $n^*(x)$, there exists a universal scaling that gives identical transmittance and reflectance in these corresponding systems. Interestingly, these scaled transmittance and reflectance reduce to their standard, non-scaled definitions in a time-reversal invariant system, i.e., one without gain or loss.

This paper is organized as follows: in Section \ref{sec:smatrix} we review the scattering matrix formalism and establish the universal scaling of transmission and reflection that becomes identical for time-reversed partners. We also provide some important identities about the scaled transmittance and reflectance. In Section \ref{sec:examples} we exemplify these result and show their connection with the physical reflectance and transmittance, through some insightful discussions that range from the Hermitian limit to parity-time ($\pt$) symmetric systems. We also relate our analysis to a special pair of time-reversed partners, i.e., a laser and a coherent perfect absorber. Finally, we provide some concluding remarks in Section \ref{sec:conclusions}, including its extension to higher dimensions.

\section{The scattering matrix} \label{sec:smatrix}

\begin{figure}[b]
	\centering
	\includegraphics[width=\linewidth]{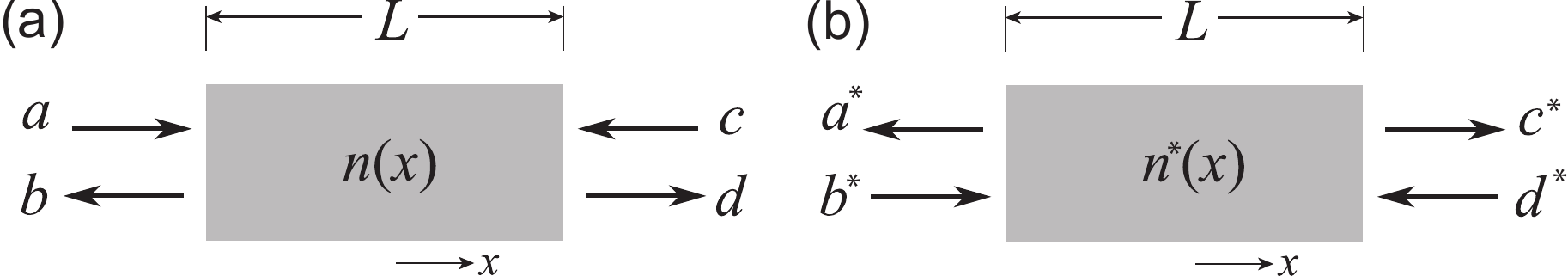}
	\caption{Two-channel scattering of light from a 1D optical system (left) and its time-reversed partner (right).}
	\label{diagram1}
\end{figure}

Consider the 1D optical system depicted in Fig.~\ref{diagram1}(a). We assume that electromagnetic waves propagate freely in space until it scatters off a cavity of a finite length $L$. The refractive index landscape in the scattering region is described by the function $n(x)$.
The transverse electric field $E(x,t)$ in the scattering region satisfies the wave equation:
\begin{equation}
	\left[\partial_x^2-\frac{n^2(x)}{c^2}\partial^2_t\right]E(x,t)=0,
	\label{waveeq}
\end{equation}
where $c$ is the speed of light in vacuum. Outside the scattering region the electric field takes the following form:
\begin{eqnarray}
	E(x,t)=\begin{cases}
	(a e^{ikx}+b e^{-ikx})e^{-i\omega t}, & x<-L/2\\
	(c e^{-ikx}+d e^{ikx})e^{-i\omega t}, & x>-L/2
	\end{cases},
	\label{eq0}
\end{eqnarray}
where $\omega$ is the real-valued frequency and $k=\omega/c$ is the wave vector in free space. The amplitudes $a$, $b$, $c$ and $d$ in Eq.~(\ref{eq0}) are depicted in Fig.~\ref{diagram1}(a), and they are related through the scattering matrix $S$ as
\begin{equation}
	\begin{pmatrix}
	b \\
	d
	\end{pmatrix}
    =\begin{bmatrix}
	  r_L(\omega) & t(\omega) \\
	  t(\omega) & r_R(\omega)
	\end{bmatrix}
    \begin{pmatrix}
	a \\
	c
	\end{pmatrix} \equiv
	S(\omega)\begin{pmatrix}
	a \\
	c
	\end{pmatrix}.
	\label{eqS0}
\end{equation}
Here $r_L$, $r_R$ are the transmission coefficients from the left and right side, and $t$ is the reciprocal transmission coefficient. As the wave equation \eqref{waveeq} is unchanged when the time $t$ is replaced by $-t$, then
\begin{eqnarray}
E(x,-t)=\begin{cases}
(a e^{ikx}+b e^{-ikx})e^{i\omega t}, & x<-L/2\\
(c e^{-ikx}+d e^{ikx})e^{i\omega t}, & x>-L/2
\end{cases}
\label{eq1}
\end{eqnarray}
is also a valid solution to the wave equation.

Next we take the complex conjugation of the wave equation \eqref{waveeq}, which now describes the scattering of light from a medium with refractive index $n^*(x)$, with the loss and gain regions exchanged from the original non-Hermitian system. Therefore, these two systems are time-reversed partners, and the electric field given by
\begin{eqnarray}
\hspace{-3mm}E^*(x,-t)=\begin{cases}
(b^* e^{ikx}+a^* e^{-ikx})e^{-i\omega t}, & x<-L/2\\
(d^* e^{-ikx}+c^* e^{ikx})e^{-i\omega t}, & x>-L/2
\end{cases}
\label{eq2}
\end{eqnarray}
indicates that the incoming amplitudes are now given by $b^*$ and $d^*$ and that they are scattered into outgoing amplitudes $a^*$ and $c^*$ [see Fig.~\ref{diagram1}(b)]. We denote the corresponding scattering matrix by $\tilde{S}$:
\be
	\begin{pmatrix}
	a^* \\
	c^*
	\end{pmatrix}=\tilde{S}(\omega)\begin{pmatrix}
	b^* \\
	d^*
	\end{pmatrix}
    =\begin{bmatrix}
	  \tilde{r}_L(\omega) & \tilde{t}(\omega) \\
	  \tilde{t}(\omega) & \tilde{r}_R(\omega)
	\end{bmatrix}
    \begin{pmatrix}
	b^* \\
	d^*
	\end{pmatrix},
\ee
or
\be
\begin{pmatrix}
	a \\
	c
	\end{pmatrix}=\tilde{S}^*(\omega)
    \begin{pmatrix}
	b \\
	d
	\end{pmatrix}
	\label{eqS1}.
\ee
By multiplying $\tilde{S}^*(\omega)$ to both sides of Eq.~(\ref{eqS0}) from the left and simplifying the result using Eq.~\eqref{eqS1}, we obtain
\begin{equation}
	\tilde{S}^*(\omega)S(\omega)=\bm{1}\label{S_identity}
\end{equation}
which relates the scattering matrix $S(\omega)$ of the original non-Hermitian system and the scattering matrix $\tilde{S}(\omega)$ of its time-reversed partner. Here $\bm{1}$ is the identity matrix.


When the determinant of $S(\omega)$ is non-zero, i.e., away from a zero of the $S$ matrix, we employ
\begin{equation}
	S^{-1}=\frac{1}{r_Lr_R-t^2}\begin{pmatrix}
	r_R & -t\\
	-t & r_L
	\end{pmatrix}
    =\tilde{S}^*
\end{equation}
to derive
\be
	\tilde{r}^*_{L,R}=\frac{r_{R,L}}{r_Lr_R-t^2},\quad	\tilde{t}^*=\frac{-t}{r_Lr_R-t^2}.
	\label{rrt1}
\ee
Furthermore, using the property that
\be
\text{det} (\tilde{S}^*S)=\text{det} \tilde{S}^* \text{det} S = 1, \label{eq:detS0}
\ee
or more explicitly,
\be
(r_Lr_R-t^2)(\tilde{r}^*_L\tilde{r}^*_R-(\tilde{t}^*)^2) = 1,\label{eq:detS}
\ee
we can rewrite Eq.~\eqref{rrt1} as
\begin{align}
	\frac{R_{R,L}}{|r_Lr_R-t^2|}&=\frac{\tilde{R}_{L,R}}{|\tilde{r}_L\tilde{r}_R-\tilde{t}^2|},\label{identity1a}\\
	\frac{T}{|r_Lr_R-t^2|}&=\frac{\tilde{T}}{|\tilde{r}_L\tilde{r}_R-\tilde{t}^2|},
	\label{identity1b}
\end{align}
where $T=|t|^2, R_{L,R} = |r_{L,R}|^2, \tilde{T}=|\tilde{t}|^2, \tilde{R}_{L,R} = |\tilde{r}_{L,R}|^2$ are the transmittance and reflectances in the two systems. Note that the subindices of the reflectances are switched in Eq.~\eqref{identity1a}, e.g., $R_R$ is related to $\tilde{R}_L$. These two relations show that there exists a universal scaling of transmittance and reflectance that is invariant after time-reversal, which utilizes the determinant of the respective scattering matrix.


\begin{figure}[h]
	\centering
	\includegraphics[width=\linewidth]{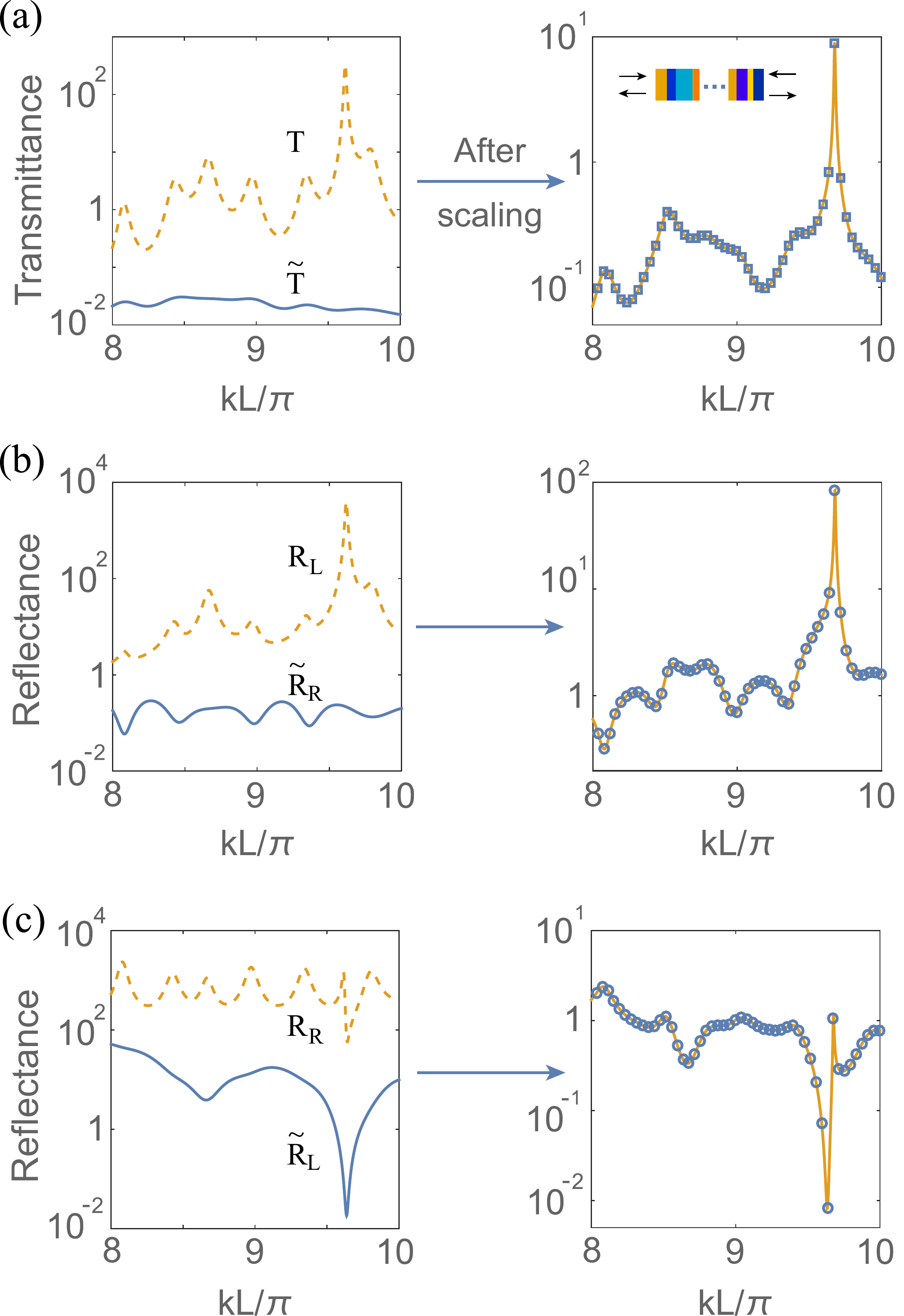}
	\caption{Scattering from a 1D heterostructure and its time-reversed partner. There are 20 layers in each structure, and the refractive index is randomly set using Re$[n]\in[1.5,2.5]$ and Im$[n]\in[-0.05,0.05]$. Layer thickness is also chosen randomly, with the widest twice as wide as the thinnest. (a) Transmittance $T$ in this system and $\tilde{T}$ in its time-reversed partner as a function of the wave number before (left) and after (right) the time-reversal invariant scaling. The line and squares represent the scaled $T$ and $\tilde{T}$, respectively. Inset: Schematic of the heterostructure. (b,c) The same for the reflectances. The lines and circles represent scaled $R_{L,R}$ and $\tilde{R}_{R,L}$, respectively.}\label{fig_random}
\end{figure}

This observation holds for 1D systems regardless of their symmetries and other properties of the refractive index. One representative example is shown in Fig.~\ref{fig_random} for a heterostructure with randomly chosen refractive indices and its time-reversed partner. Not only does their transmittance (as well as reflectances) differ as much as four orders of magnitude, but the frequency-dependence of these spectra also shows distinct features. Nevertheless, once scaled by the universal factor given in Eqs.~\eqref{identity1a} and \eqref{identity1b}, their transmittances and corresponding reflectances become identical, respectively.

The time-reversal invariant scaling can also be derived using the $R$-matrix construction of the $S$ matrix \cite{CPA,wigner_eisenbud_1947}:
\be
  S = - e^{2in_0kx_0}\, \left[\bm{1} - in_0kR \right]^{-1} \, \left[\bm{1} + in_0kR \right],
\ee
where $x_0 > L/2$ is an arbitrary boundary and $n_0$ is the real-valued refractive index in free space. To obtain $\tilde{S}$ for the time-reversed partner system, we use the same expression and utilize the fact that the $R$ matrix becomes its complex conjugate when $n(x)$ is replaced by $n^*(x)$. As a result, we find
\be
  \tilde{S}^* = - e^{-2in_0kx_0}\, \left[\bm{1} + in_0kR \right]^{-1} \, \left[\bm{1} - in_0kR \right],
\ee
which leads to Eq.~\eqref{S_identity} and in turn Eqs.~\eqref{identity1a} and \eqref{identity1b}.

As we work through some enlightening examples we will unravel the physical significance of these and other scaled reflectances and transmittances in systems obeying particular symmetries, as described in the following sections.

\section{Examples} \label{sec:examples}

\subsection{Systems with time-reversal symmetry}

We start our exemplification by considering a dielectric slab with an arbitrary index profile $n(x)$, where the only constraint is that $n(x)$ is real, i.e., the slab is dissipationless and gainless. Then the slab is its own time-reversed partner, and in this case it is interesting to note that the scaled transmittance and reflectance in Eqs.~\eqref{identity1a} and \eqref{identity1b} reduce to their standard and unscaled forms. This is because now $\tilde{S}$ is the same as $S$, leading to $|\text{det} S|=|r_Lr_R-t^2|=1$ in Eq.~\eqref{eq:detS}, i.e., the time-reversal invariant scaling vanishes in this case.

However, we should note that
\be
\frac{T}{|r_Lr_R-t^2|}+\frac{R_{R,L}}{|r_Lr_R-t^2|}\neq1
\ee
in general, and hence one may wonder what the general ``conservation law" is that reduces to $T+R=1$ in the time-reversal invariant case, where $R_L=R_R\equiv R$.
It turns out that such a ``conservation law" is simply given by
\be
\mathfrak{T}+\mathfrak{R}=1,\label{eq:trivial_law}
\ee
where we have defined
\be
	\mathfrak{R}\equiv\frac{r_Lr_R}{r_Lr_R-t^2},\quad	\mathfrak{T}\equiv\frac{-t^2}{r_Lr_R-t^2}
	\label{pseudoRT}
\ee
as the pseudo-reflectance and pseudo-transmittance.

$\mathfrak{R}$ and $\mathfrak{T}$ are complex in general, but they also become the physical reflectance and transmittance in the time-reversal symmetric case. To show explicitly that $\mathfrak{R}$ and $\mathfrak{T}$ are real in this case, we note that Eq.~\eqref{rrt1} now becomes
\be
	r^*_{L,R}=\frac{r_{R,L}}{r_Lr_R-t^2},\quad	t^*=\frac{-t}{r_Lr_R-t^2}.
\ee
Therefore, we find
\begin{gather}
|r_L|^2=|r_R|^2=\frac{r_Lr_R}{r_Lr_R-t^2}=\mathfrak{R},\\
|t|^2 = \frac{t^2}{r_Lr_R-t^2}=\mathfrak{T},
\end{gather}
and the trivial ``conservation law" given by Eq.~\eqref{eq:trivial_law} now becomes the actual flux conservation relation. The latter, of course, can also be obtained from the unitarity property of the scattering matrix, i.e., $S^\dagger S=1$, which can be derived using Eq.~(\ref{S_identity}) with the assumed reciprocity ($S^T=S$) in this case.

\subsection{A dielectric slab with gain or loss}

One of the simplest systems that exhibit nontrivial scattering features is a 1D slab which is capable of amplify or absorb radiation. The former is a typical model used to study solid-state laser cavities, and the consideration of its time-reversed partners has led to the discovery of coherent perfect absorbers (CPA) \cite{longhiCPA,CPA,CPA_exp}.

However, such a consideration has only been explored for the extremes of the scattering matrix, namely its poles and zeros. A pole (zero) of a scattering matrix is defined as the usually complex frequency where one or more of eigenvalues of the scattering matrix become infinite (zero). Therefore, at a pole of a scattering matrix an infinitesimal input (e.g., noises as input $a$ and $c$ in Fig.~\ref{diagram1}) can lead to a finite scattering or output light intensity, which is one mathematical model used to describe a laser with a gain medium when the pole occurs on the real frequency axis. Clearly, a ``time-reversed laser" then corresponds to a zero of the scattering matrix, where the incoming light satisfying certain coherent phase and amplitude configuration is absorbed perfectly by the time-reversed system with loss, after which the CPA is named.

\begin{figure}[b]
	\centering
	\includegraphics[width=\linewidth]{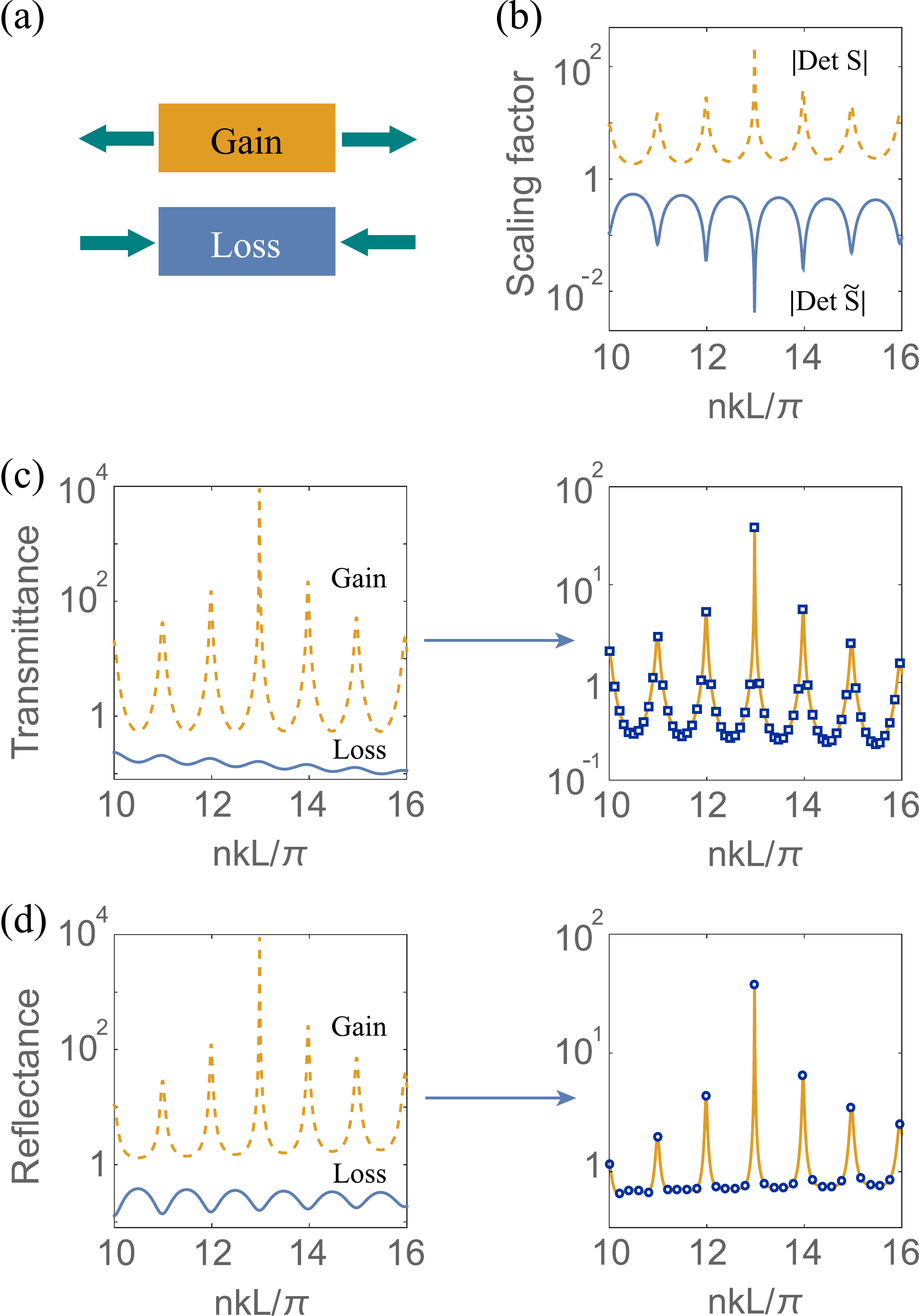}
	\caption{Scattering from a pair of uniform dielectric slabs with gain and loss, respectively. $n=3\pm0.05i$. (a) Schematics showing their respective lasing and CPA state. (b) Scaling factors as a function of the wave number. (c) Transmittance before (left) and after (right) the time-reversal invariant scaling. (d) Same for the reflectance. The symbols are the same as in Fig.~\ref{fig_random}.}
\label{fig_cpa}
\end{figure}

Such a correspondence is a special case of Eq.~\eqref{eq:detS0}: a pole (zero) implies that the determinant of the scattering matrix is infinite (zero), and the time-reversal relation between a laser and a CPA is described asymptotically by Eq.~\eqref{eq:detS0}: $\text{det}S\rightarrow\infty$ in a laser cavity with gain while $\text{det}\tilde{S}\rightarrow 0$ in the corresponding CPA with loss [see Fig.~\ref{fig_cpa}(b)].

It is well known that $\text{det} S = r_Lr_R-t^2\rightarrow \infty$ implies that $r_L,r_R$ and $t$ all diverge at the frequency of a lasing mode. However, one usually cannot verify analytically whether these quantities approach infinity at the same speed without knowing their explicit expressions. Together with Eq.~\eqref{eq:detS}, the scaling relation \eqref{rrt1} offers us a unique opportunity to overcome this difficulty. Specifically, they lead to $\mathfrak{\tilde{R}}^*=\mathfrak{R}$ for the pseudo-reflectance in a pair of time-reserved partner systems, or more explicitly,
\be
\left[\frac{\tilde{r}_L\tilde{r}_R}{\tilde{r}_L\tilde{r}_R-\tilde{t}^2}\right]^* = \frac{r_Lr_R}{r_Lr_R-t^2}.\label{eq:asymptotic}
\ee
As $\text{det} S = r_Lr_R-t^2\rightarrow \infty$ in a laser cavity, $\text{det} \tilde{S} = \tilde{r}_L\tilde{r}_R-\tilde{t}^2\rightarrow 0$ in the corresponding CPA as we have mentioned, and $\tilde{r}_L\tilde{r}_R$ remains finite in this process. As a result, the left hand side of Eq.~\eqref{eq:asymptotic} diverges, and so does its right hand side. Because the denominator of the latter (i.e., $\text{det} S$) also diverges, it indicates that $r_Lr_R$ approaches infinity faster than $r_Lr_R-t^2$. In other words, the leading order asymptotics of $r_Lr_R$ and $t^2$ are the same. In a system with reflection symmetry as the one shown in Fig.~\ref{fig_cpa}, $r_L=r_R$ and hence they approach infinity at the same speed as $t$ [see the highest peak around $10^4$ at $nkL/\pi \approx 13$ in Figs.~\ref{fig_cpa}(c) and (d)].

Similar to what we have seen in Fig.~\ref{fig_random}, here the physical reflectance and transmittance differ significantly in the gain and loss slabs [see the left panels in Figs.~\ref{fig_cpa}(c) and (d)]. After applying the scaling factor specified in Eqs.~\eqref{identity1a},\eqref{identity1b} and shown in Fig.~\ref{fig_cpa}(b), we again verify the time-reversal invariant scaling of the reflectance and transmittance [see the right panels in Figs.~\ref{fig_cpa}(c) and (d)].

\subsection{$\pt$-symmetric systems}

Similar to how we derived Eq.~\eqref{S_identity} for a pair of time-reversal partners, there is an identity that relates the scattering matrix $S$ of a medium with refractive index $n(x)$ and the scattering matrix $\bar{S}$ of the $\pt$-symmetric partner with refractive index $\bar{n}(x)=n^*(-x)$:
\be
\sigma_x\bar{S}^*(\omega)\sigma_x S(\omega)=\bm{1}.
\ee
Here $\sigma_x$ is the first Pauli matrix, and the counterpart to Eq.~\eqref{rrt1} is
\be
	\bar{r}^*_{L,R}=\frac{r_{L,R}}{r_Lr_R-t^2},\quad	\bar{t}^*=\frac{-t}{r_Lr_R-t^2}.
	\label{rrt_pt}
\ee
Here the additional parity operator simply leaves the reciprocal transmission unchanged (i.e., $\bar{t}=\tilde{t}$) and exchanges the left and right reflection coefficients (i.e., $\bar{r}_{L,R}=\tilde{r}_{R,L}$). Consequently, the time-reversal invariant scaling also indicates the following $\pt$-invariant scaling:
\begin{align}
	\frac{R_{L,R}}{|r_Lr_R-t^2|}&=\frac{\bar{R}_{L,R}}{|\bar{r}_L\bar{r}_R-\bar{t}^2|},\\
	\frac{T}{|r_Lr_R-t^2|}&=\frac{\bar{T}}{|\bar{r}_L\bar{r}_R-\bar{t}^2|}.
\end{align}

If a non-Hermitian system is $\pt$-symmetric, then we can write its physical transmittance and reflectances as
\be
	R_{L,R}=\frac{r_{L,R}^2}{r_Lr_R-t^2}\in\mathbb{R},\quad 	
    T=\frac{-t^2}{r_Lr_r-t^2}\in\mathbb{R}
\ee
using Eq.~\eqref{rrt_pt}. The generalized conservation law \cite{Ge2012} then follows as a consequence, i.e.,
\begin{equation}
	|1-T|=\sqrt{R_LR_R}.
\end{equation}




\section{Conclusion and Discussions} \label{sec:conclusions}

In summary, we have shown that there exists a time-reversal invariant scaling for wave propagation in 1D non-Hermitian systems. It applies to both transmission and reflection, no matter how different in magnitude and frequency-dependency the spectra of these quantities are in this system and its time-reversed partner.

Although we have restricted our discussion to 1D systems so far, some of our observations can be easily extended to higher dimensions. For example, for a quasi-1D waveguide with multiple transverse channels \cite{ge_scattering_2015}, the relation (\ref{S_identity}) still applies as long as the incoming and outgoing channels with the same index are related by time reversal. In this case the scattering matrices can be written in their block forms, i.e.,
\be
S(\omega) =
\begin{pmatrix}
\bm{r}_L & \bm{t}\\
\bm{t} & \bm{r}_R
\end{pmatrix},\quad
\tilde{S}(\omega) =
\begin{pmatrix}
\tilde{\bm{r}}_L & \tilde{\bm{t}}\\
\tilde{\bm{t}} & \tilde{\bm{r}}_R,
\end{pmatrix}
\ee
where $\bm{r}_{L,R},\bm{t},\tilde{\bm{r}}_{L,R},\tilde{\bm{t}}$ become matrices themselves. We then find the following identity:
\be
\underline{T}+\underline{R}=N,
\ee
where $\underline{T}\equiv\tilde{\bm{t}}^*\bm{t}$, $\underline{R}\equiv\text{Tr}(\tilde{\bm{r}}_L^*\bm{r}_L) = \text{Tr}(\tilde{\bm{r}}_R^*\bm{r}_R)$, and $N$ is the number of incoming (and outgoing) channels.

Similarly, the identity (\ref{S_identity}) also holds for the scattering of cylindrical waves in two dimensions (2D). Following the convention used in Ref.~\cite{ge_constructing_2017}, we define the $m$th incoming and outgoing channels by
\be
\Psi_m^-(r,\theta) = \frac{H^-_m(kr)}{H^-_m(kR)}e^{im\theta},\quad \Psi_m^+(r,\theta) = \frac{H^+_m(kr)}{H^+_m(kR)}e^{im\theta}.\nonumber
\ee
Here $m\in\mathbb{Z}$ is the angular momentum number, and $m>0$ ($m<0$) describes counterclockwise (clockwise) waves. $r,\theta$ are the radial position and the azimuthal angle, $R$ is the radius of the scattering region, and $H^\pm$ are the Bessel functions of the first and second kind. In the absence of scattering, the scattering matrix becomes an anti-diagonal matrix, similar to the 1D case defined in Eq.~\eqref{eqS0}.

\begin{figure}[b]
	\centering
	\includegraphics[width=\linewidth]{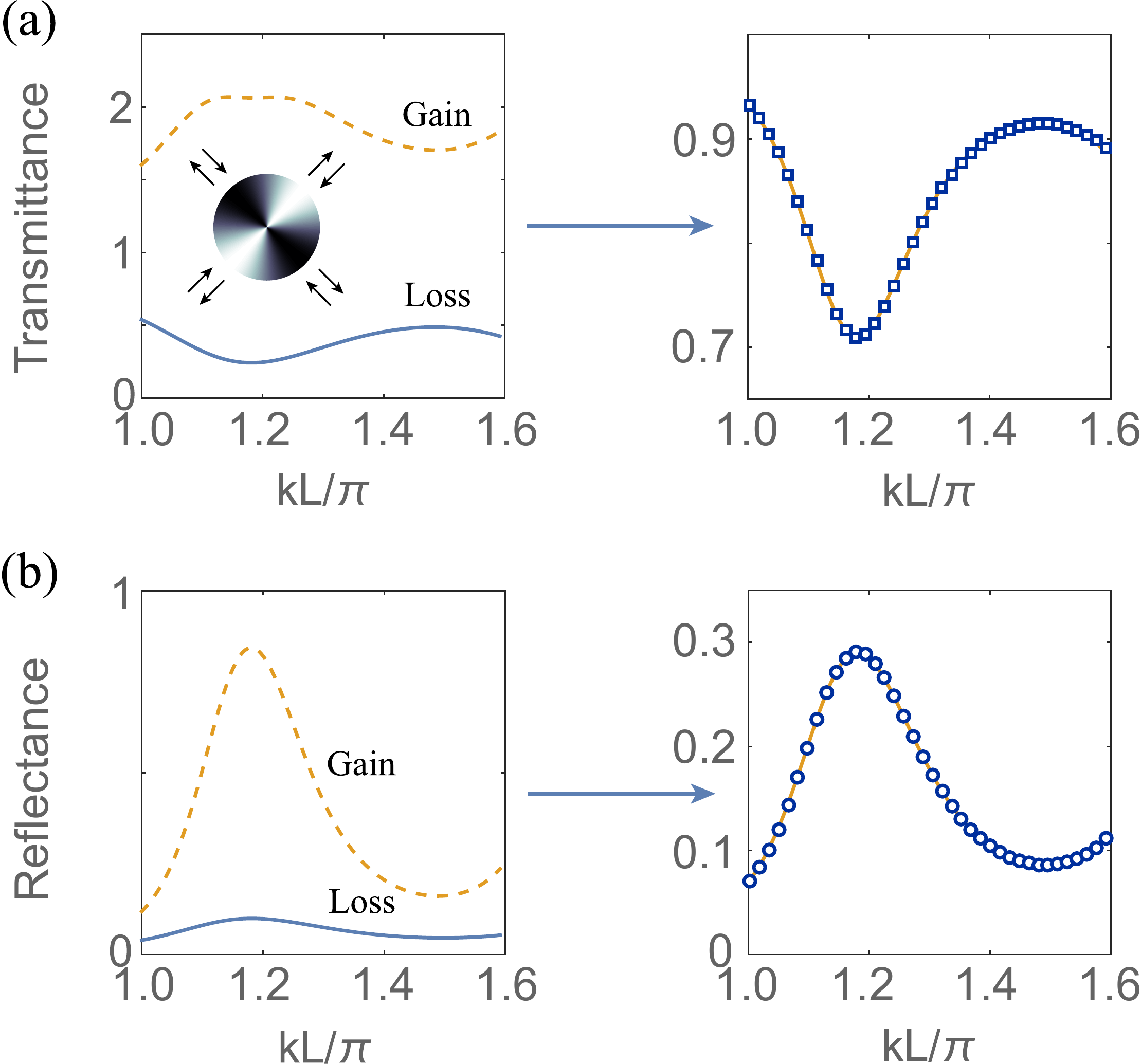}
	\caption{Time-reversal invariant scaling of a reduced scattering matrix in 2D disk geometry. (a) Transmittance as a function of the wave number before (left) and after (right) the time-reversal invariant scaling. Inset: Schematic of the disk scatterer. (b) The same for the reflectance. The symbols are the same as in Fig.~\ref{fig_random}.}\label{fig_2D}
\end{figure}

Although the time-invariant scaling of the scattering coefficients does not exist in higher-dimensions in general, it reemerges in some special cases. Take the scattering of the cylindrical waves, for example. If the index modulation itself has an angular momentum $M=2$ (e.g., a quadruple cavity with a uniform refractive index \cite{song_channeling_2012,liu_transporting_2018}), then a cylindrical wave of angular momentum $m$ will be scattered strongly into the $m\pm2$ channels in general \cite{ge_extreme_2013}. More specifically, the $m=\pm1$ channels are scattered strongly into each other and the $m'=\pm3$ channels. If the index modulation is weak, then the scattering intensities into the $m'=\pm3$ channels are negligible due to a low spectral overlap factor \cite{ge_controlling_2013}. Consequently, the $m=\pm1$ channels form a largely closed subspace of the scattering matrix, and the reduced-dimension scattering matrix $S$ in these channels behave in the same way as in the 1D case.

Assuming reciprocity, we denote the scattering amplitudes between the $m=\pm1$ channels as the ``transmission coefficient" $t$ and those back into the outgoing channels of the same indices as the reflection coefficients $r_{L,R}$. The same time-reversal invariant scaling shown in Eqs.~\eqref{identity1a} and \eqref{identity1b} still holds, and we exemplify this result using two disks of radius $R$ and refractive index $n(r,\theta) = (1.5+0.1\sin\,2\theta)\pm0.05i$. To verify that the reduction of the scattering matrix into the $2\times 2$ form is a good approximation, we first mention that the ratio of the scattering intensities into the $m'=3$ and $m'=1$ ($-1$) channels from the $m=1$ channel is $1.3\times10^{-2}$ ($6.1\times10^{-4}$) at $kR=3$, which is typical for the range of wave numbers shown in Fig.~\ref{fig_2D}. The reflectances $R_{R,L}$ in this case are identical, because the $m=\pm 1$ channels are exchanged when we simply change our perspective from the top view to the bottom view of the 2D plane. In other words, the chirality of the channels is flipped when $\theta\rightarrow-\theta$. The transmittances $T$ and $\tilde{T}$ shown in Fig.~\ref{fig_2D} do not exhibit similarities in particular, but they become identical after applying the time-reversal invariant scaling specified in Eq.~(\ref{identity1b}).

This work is supported by NSF under Grant No. DMR-1506987 and by PSC-CUNY under Award No. 61787-49.

\bibliography{references}

\end{document}